\documentclass{ws-procs975x65}
\usepackage{cite}
\usepackage{url}

\begin{document}

\title{PRODUCTION AND EVAPORATION OF PLANCK SCALE BLACK HOLES AT THE LHC}
\author{P. NICOLINI$^{(a)}$, J. MUREIKA$^{(b)}$, E. SPALLUCCI$^{(c)}$, E. WINSTANLEY$^{(d)}$ and M. BLEICHER$^{(a)}$}
\address{$^{(a)}$Frankfurt Institute for Advanced Studies and Institut f\"{u}r Theoretische Physik,\\
Johann Wolfgang Goethe-Universit\"{a}t,\\
Frankfurt am Main, D-60348, Germany\\
E-mail: nicolini@th.physik.uni-frankfurt.de, bleicher@th.physik.uni-frankfurt.de
}
\address{$^{(b)}$Department of Physics, Loyola Marymount University,\\
Los Angeles, CA 90045, USA\\
E-mail: jmureika@lmu.edu}
\address{$^{(c)}$Dipartimento di Fisica Teorica, Universit\`a di Trieste
and INFN, Sezione di Trieste, \\ Trieste, I-34151,  Italy\\E-mail: spallucci@ts.infn.it}
\address{$^{(d)}$Consortium for Fundamental Physics, School of Mathematics and Statistics,
The University of Sheffield,\\ Sheffield, S3 7RH, United Kingdom\\
E-mail: E.Winstanley@sheffield.ac.uk}
\begin{abstract}
We review the phenomenology of mini black holes at colliders in light of the latest data from the LHC. By improving the conventional production cross-section, we show that the current non-observation of black hole signals can be explained in terms of quantum gravity effects.  In the most optimistic case, black hole production could take place at a scale slightly above the LHC design energy. We also analyse possible new signatures of quantum-corrected Planck-scale black holes: in contrast to the semiclassical scenario the emission would take place in terms of soft particles mostly on the brane.
\end{abstract}
%

 \bodymatter\bigskip

The main argument in support of mini black hole (BH) production\cite{bhlhc} is based on a viable solution of the hierarchy problem of particle physics: by allowing gravity to probe additional spatial dimensions, one can  avoid the discrepancy between the electroweak scale and the Planck scale, placing quantum gravity phenomenology in reach of current particle accelerators\cite{reviews}. Such logic, however, is being questioned by the current non-observation of quantum gravity signals at the LHC\cite{expdata}, 
giving the topic of mini BHs a more speculative character with respect to early expectations 
(see Ref. \refcite{qg} for recent reviews of the status of quantum gravity and innovative proposals to test it).

Against this background, we show that
the mechanism of BH production and their associated signatures might be drastically different from what  is known to date.  Our general argument is based on the fact that semiclassical formulations cannot efficiently describe the conjectured quantum gravity at the terascale.
Even if we acknowledge the difficulties in predicting quantum gravity phenomenology from first principles (\textit{e.g.} from string theory, loop quantum gravity, {\it etc...}), we recall that irrespective of the specific formulation, quantum gravity exhibits a unique character: the emergence of a minimal resolution length $\ell$. As a result, one can follow the strategy of improving the existing scenarios by  
any of the minimal length mechanisms proposed in the literature (for recent reviews see Ref. \refcite{ml}).
%
%

Customarily, mini-BH formation is modelled via Thorne's ``hoop conjecture'': a BH forms whenever the
impact parameter $b$ becomes smaller than the effective Schwarzschild
radius, $r_\mathrm{H}$, of the two-body colliding
system. While for classical BHs such a conjecture consistently implies a
black disk  cross section, $\sigma_\mathrm{BH}=\pi r_\mathrm{H}^2$, in
the case of particle collisions quantum effects are expected to be
relevant.  Indeed, the black disk profile leads to the
phenomenologically inconsistent result that about $\sim 1\
\mathrm{BH}/\mathrm{day}$ would have formed
 at the Super Proton Synchrotron (SPS) in 1985, provided additional spatial dimensions are assumed. To overcome this oddity, we recall that even the impact parameter cannot be smaller than the minimal  length, $\ell\lesssim b$.
%
%
%
%
%
One can thus analytically improve the BH cross section as\cite{MNS12}
\begin{eqnarray}
\sigma_\mathrm{BH}(s)=\pi\ell^2\Gamma\left(-1; \ell^2/r_\mathrm{H}^2(s)\right), \quad
 \Gamma\left(\alpha; x\right)=\int_x^\infty dt \, t^{\alpha-1}e^{-t}, \label{cs}
\end{eqnarray}
where $\sqrt{s}$ is the Mandelstam ``invariant mass'' of the colliding two-body system.  The above formula smoothly interpolates the black disk result in the trans-Planckian regime and the expected suppression of sub-Planckian  BHs, without introducing ``by hand'' any threshold function. By setting $\ell\sim 10^{-19}\ \mathrm{m}$ and considering the current limits on the LHC luminosity\cite{LHC13} $\sim 7.73\times 10^{37}\ \mathrm{m}^{-2} \mathrm{s}^{-1}$, we can estimate the BH production rate $\dot{N}_\mathrm{BH}$. In doing so, one has also to take into account further quantum corrections to $r_\mathrm{H}(s)$, which departs from the classical linear function of the beam energy $\sqrt{s}$ due to BH remnant formation\cite{remnant}.  In conclusion, one has a variety of parameter combinations but $\dot{N}_\mathrm{BH}$ turns out to be negligible except for a high number $n$ of extra dimensions. For M-theory inspired $11$-dimensional spacetimes, the ``new physics'' might be just behind the corner, \textit{i.e.},  $\dot{N}_\mathrm{BH}\sim 6/\mathrm{year}$ for $\sqrt{s}\sim 16$ TeV. We notice that the resulting suppression \textit{is} a genuine effect of quantum gravity while the semiclassical regime occurs at energies well above the terascale.

Another key point that needs revision is the profile of BH emission spectra.  Conventionally these are obtained by considering greybody factors resulting from scattering equations on classical higher dimensional BH geometries. This analysis cannot work for Planck scale black holes (QBHs): strong quantum effects at scales $\sim \ell$ not only smear out the curvature singularity but modify the global structure of the geometry with consequent horizon extremisation and  halted evaporation
\cite{qgbhs}. 
At the level of primary emission, QBHs exhibit a decreased temperature with respect to classical BHs with the same mass: this implies reduced fluxes of particles and energy both in the brane and in the bulk. Such a reduction, however, is not ``homogeneous'': in marked contrast to the results for Schwarzschild BHs\cite{HaK03}, the ratio of total bulk/brane emission can be seen to decrease rapidly as $n$ increases\cite{NiW11} (see Tab.~\ref{aba:tbl1}). 
\begin{table}
\tbl{Relative bulk-to-brane energy emission rates for scalar fields.\cite{NiW11}}
{\begin{tabular}{@{}lcccccccc@{}}
\toprule
 & $n=0$ & $n=1$ &$n=2$ &$n=3$ &$n=4$ &$n=5$ &$n=6$ &$n=7$
\\\colrule
Schwarzschild BH\hphantom{00} & $1.0$ & $0.40$ & $0.24$\hphantom{0} & $0.22$\hphantom{0} & $0.24$\hphantom{00} & $0.33$\hphantom{00} & $0.52$\hphantom{000} & $0.93$\hphantom{000} \\
QBHs& $1.0$ & $0.27$ & $0.082$ & $0.027$ & $0.0089$ & $0.0029$ & $0.00095$ & $0.00028$\\\botrule
\end{tabular}}
\label{aba:tbl1}
\end{table}
Even if our conclusions are based on the primary emission of scalar particles only and a study of the dynamics of the associated chromosphere and photosphere is still missing, we argue that QBHs can be characterized by a distinctive signature, \textit{i.e.}, a peculiar emission of detectable soft particles mostly on the brane.
%
%
%
%
%
%
%
%
%
%
%
%
%
%
%
%
%
%
%
%
%
%
%
\section*{Acknowledgments}
This work has been supported by the German Research Foundation (DFG) grant NI 1282/2-1,
 by the Helmholtz International Center for FAIR within the framework of the LOEWE program (Landesoffensive zur Entwicklung Wissenschaftlich-\"{O}konomischer Exzellenz) launched by the State of Hesse, by a Continuing Faculty Grant from Loyola Marymount University, by the Lancaster-Manchester-Sheffield Consortium for Fundamental Physics under STFC grant ST/J000418/1 and partially 
 by the European 
COST action MP0905 ``Black Holes in a Violent Universe''.
%
%
%
%
%

\end{document}